\documentclass[10pt,conference]{IEEEtran}

\usepackage{url,amsmath,amssymb,color}
\usepackage[T1]{fontenc}
\renewcommand{\Pr}{\mbox{{\rm Pr}}}
\newcommand{\E}{{\cal E}}
\newcommand{\bx}{{\bf x}}

\newcommand{\bs}{{\bf s}}
\newtheorem{definition}{Definition}
\newtheorem{proposition}[definition]{Proposition}
\newtheorem{corollary}[definition]{Corollary}
\newtheorem{theorem}[definition]{Theorem}

\newenvironment{example}{\begin{trivlist}\item[]{\bf
      Example.}}{\end{trivlist}}

\begin{document}

\title{Computing the biases of parity-check relations}
\author{
\IEEEauthorblockN{Anne Canteaut}
\IEEEauthorblockA{INRIA project-team SECRET\\
B.P. 105 \\78153 Le Chesnay Cedex, France\\
Email: Anne.Canteaut@inria.fr}
\and
\IEEEauthorblockN{María Naya-Plasencia}
\IEEEauthorblockA{INRIA project-team SECRET\\
B.P. 105 \\78153 Le Chesnay Cedex, France\\
Email: Maria.Naya\_Plasencia@inria.fr}}
\maketitle

\begin{abstract}
A divide-and-conquer cryptanalysis can often be mounted against some keystream generators composed of several (nonlinear) independent devices combined by a Boolean function. In particular, any parity-check relation derived from the periods of some constituent sequences usually leads to a distinguishing attack whose complexity is determined by the bias of the relation. However, estimating this bias is a difficult problem since the piling-up lemma cannot be used. Here, we give two exact expressions for this bias. Most notably, these expressions lead to a new algorithm for computing the bias of a parity-check relation, and they also provide some simple formulae for this bias in some particular cases which are commonly used in cryptography.
\end{abstract}

\section{Divide-and-conquer attacks against some stream ciphers}
\begin{figure}[h]
\begin{center}
\setlength{\unitlength}{0.005in}%
\begin{picture}(310,170)(40,605)
\thicklines
\put( 20,620){\framebox(120,30){Device \(n\)}}
\put( 20,690){\framebox(120,30){Device 2}}
\put( 20,740){\framebox(120,30){Device 1}}

\put(100,658){\makebox{\vdots}}

\put(220,660){\framebox(60,60){\(f\)}}

\put(140,750){\line( 1, 0){ 40}}
\put(180,750){\vector( 1,-1){ 38}}
\put(140,630){\line( 1, 0){ 40}}
\put(180,630){\vector( 1, 1){ 38}}

\put(140,700){\vector( 1, 0){ 80}}
\put(280,690){\vector( 1, 0){ 40}}

\put(330,685){\makebox(0,0)[lb]{\(\bs\) keystream}}
\put(200,740){\makebox(0,0)[lb]{\(\bx_1\)}}
\put(185,675){\makebox(0,0)[lb]{\(\bx_2\)}}
\put(205,630){\makebox(0,0)[lb]{\(\bx_n\)}}
\end{picture}

\caption{\label{fig: generator}Keystream generator composed of several independent devices combined by a Boolean function}
\end{center}
\end{figure}
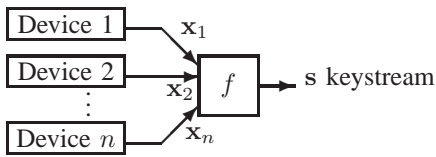

Parity-check relations are extensively used in cryptanalysis for
building statistical distinguishers. For instance, they can be exploited in
divide-and-conquer attacks against some stream ciphers which consist
of several independent devices whose output sequences are combined by
a nonlinear function. Here, we focus on such keystream generators as depicted on Figure~\ref{fig: generator}. All the \(n\)~constituent devices are updated independently from each other. The only assumption which will be used in the whole paper is that each sequence \({\bf x_i} = (x_i(t))_{t \geq 0}\) generated by the \(i\)-th device is periodic with least period~\(T_i\).

The simplest case of a generator built according to the model depicted
in Figure~\ref{fig: generator} is the combination generator, where all
devices are LFSRs. However, our work is of greater interest in the
case where the next-state functions of the constituent devices are
nonlinear. The eSTREAM candidate Achterbahn and its
variants~\cite{achterbahn,achterbahn128},
designed by Gammel, Göttfert and Kniffler, follow this design
principle: all these ciphers are actually composed of several nonlinear
feedback shift registers (NLFSRs) with maximal periods. This design is
very attractive since the use of independent devices enables to accommodate a 
large internal state with a small hardware footprint. 

However, the main weakness of this design is obviously that it is inherently vulnerable to divide-and-conquer attacks. As originally pointed out by Siegenthaler~\cite{Sie85}, the cryptanalyst may actually mount an attack which depends on a small subset of the constituent devices only. This can be done if there exists a smaller generator which involves \(k\)~constituent devices whose output is correlated to the keystream. This equivalently means that there exists a correlation between the output of the combining function and the output of a Boolean function depending on \(k\)~variables. The smallest number~\(k\) of devices that have to be considered together in the attack is then equal to \((t+1)\) where \(t\) is the correlation-immunity order (or resiliency order) of the combining function~\(f\).
Recall that a Boolean function is said to be \(t\)-th order correlation-immune if its output distribution does not change when any \(t\) input variables are fixed. Moreover, a \(t\)-resilient function is a \(t\)-th order correlation-immune function which is balanced.

Now, we recall how parity-check relations can be used for mounting a divide-and-conquer attack against such a keystream generator. This technique has been introduced by Johansson, Meier and Muller~\cite{Johansson_Meier_Muller06} for cryptanalysing the first version of Achterbahn~\cite{achterbahn}. Then, it has been extensively exploited in several attacks against the following variants of the cipher~\cite{Hell_Johansson06,Naya07,Hell_Johansson07,Naya08}.
By analogy with coding theory, a parity-check relation for a binary sequence \({\bf x}= (x(t))_{t \geq 0}\) is a linear relation between some bits of \({\bf x}\) at different instants \((t+\tau)\) where \(\tau\) varies in a fixed set and \(t\) takes any value:
\[\bigoplus_{\tau \in {\cal T}} x(t + \tau) = 0, \;\; \forall t \geq 0.\]
Then, the indexes \(\tau\) corresponding to the nonzero coefficients of the characteristic polynomial of a linear recurring sequence provide a parity-check relation. A two-term parity-check relation, 
\[x(t) \oplus x(t + \tau) = 0, \;\; \forall t \geq 0,\]
obviously corresponds to a period of the sequence. 
In the following, we only focus on parity-check relations between \(2^s\) instants which are defined as follows.
\begin{definition}
Let \(\bx_1, \ldots, \bx_n\) be \(n\) sequences and let \(f\) be a Boolean function of \(n\)~variables. Then, for any set 
\[{\cal T}=\big\{\sum_{i=1}^s c_i M_i, \;\; c_i \in \{0,1\}\big\}\]
where \(M_1, \ldots, M_s\) are some non-negative integers, 
\(PC_{f,{\cal T}}\) is the binary sequence defined by
\[PC_{f,{\cal T}}(t)  = \bigoplus_{\tau \in {\cal T}} f(x_1(t + \tau) , \ldots, x_n(t + \tau)), \; \forall t \geq 0.\]
\end{definition}

In the following, each \(M_i\) corresponds to a multiple of the least common multiple of the periods of some constituent sequences. Moreover, in order to simplify the notation, we will assume without loss of generality that the input variables are ordered in such a way that each integer~\(M_i\) corresponds to a multiple of \({\rm lcm}(T_{\ell_i+1}, \ldots, T_{\ell_{i+1}})\) with \(\ell_1 = 0\) and \(\ell_{s+1} = k\). This notably implies that \({\cal T}\) involves the periods of the first \(k\)~sequences, \(\bx_1\) \ldots, \(\bx_k\).

\begin{proposition}
Let \(\bx_1, \ldots, \bx_n\) be \(n\) sequences with least periods \(T_1, \ldots, T_n\) and 
\[{\cal T}=\big\{\sum_{i=1}^s c_i M_i, \;\; c_i \in \{0,1\}\big\}\]
where \(M_i = q_i {\rm lcm}(T_{\ell_i+1}, \ldots, T_{\ell_{i+1}})\) with \(q_i > 0\) and \(\ell_1=0\) and \(\ell_{s+1}=k\).
Let \(g\) be any Boolean function of \(k\)~variables of the form
\[g(x_1, \ldots, x_k) = \sum_{i=1}^s g_i(x_{\ell_i+1}, \ldots, x_{\ell_{i+1}})\]
where each \(g_i\) is any Boolean function of \((\ell_{i+1}-\ell_i)\)~variables. Then, for all \(t \geq 0\), we have
\[PC_{g, {\cal T}}(t) = \bigoplus_{\tau \in {\cal T}} g(x_1(t + \tau) , \ldots, x_n(t + \tau)) = 0.\]
\end{proposition}

In the whole paper, we use the following notation.
\begin{definition}
Let \(f\) be a Boolean function of \(n\)~variables. Then, the {\em bias} of \(f\) is
\[\E(f) =  2^{-n} \sum_{x \in {\bf F}_2^n} (-1)^{f(x)}.\]
This quantity is also called the imbalance of~\(f\) (e.g. in~\cite{Harpes_Kramer_Massey95,Kukorelly}) or the correlation between \(f\) and the all-zero function (e.g. in~\cite{Nyberg01}).
\end{definition}

The underlying principle of the attack presented by Johansson, Meier and Muller~\cite{Johansson_Meier_Muller06} consists in exhibiting a biased approximation~\(g\) of the combining function \(f\) which involves \(k\)~input variables, and a parity-check relation \(PC_{g, {\cal T}}=0\)  for the sequence \(g(\bx_1, \ldots, \bx_k)\). Then, the associated parity-check relation applied to \(f(\bx_1, \ldots, \bx_n)\) does not vanish but it is biased in the sense that it is not uniformly distributed when the \((T_1+ \ldots+ T_n)\) bits \(x_1(0), \ldots, x_1(T_1-1), x_2(0), \ldots, x_2(T_2-1), \ldots, x_n(T_n-1)\) are randomly chosen. The bias of \(PC_{f, {\cal T}}\), denoted by \(\E(PC_{f, {\cal T}})\) is then defined as the bias of a Boolean function with \((T_1+ \ldots+ T_n)\)~input variables corresponding to the concatenation of the first periods of the sequences.
It follows that
\[\Pr [PC_{f, {\cal T}}(t)=0] = \frac{1}{2}(1+ \E(PC_{f, {\cal T}}))\]
with \(\E(PC_{f, {\cal T}}) > 0\).
Then, computing  
\[PC_{f, {\cal T}}(t) = \bigoplus_{\tau \in {\cal T}} s(t + \tau)\]
where \(\bs\) is the keystream for different values of~\(t \geq 0\) enables the attacker to distinguish the keystream from a random sequence.
The complexity of this distinguishing attack depends on the bias \(\varepsilon\) of \(PC_{f, {\cal T}}\). More precisely, the time complexity of the attack corresponds to \(\varepsilon^{-2} 2^s\) where \(2^s\) is the number of elements in \({\cal T}\) since the bias \(\varepsilon\) can be detected from at least \(\varepsilon^{-2}\) occurrences of the biased relation. The data complexity, {\em i.e.} the number of consecutive keystream bits required for the attack is then the maximal value which must be considered for \((t + \tau)\), {\em i.e.}
\[\varepsilon^{-2} + \max {\cal T}.\]
Many variants of this attack can be derived 
\cite{Hell_Johansson06,Naya07,Hell_Johansson07,Naya08}.  However,
determining the complexity of all these attacks requires an estimation
of the bias of \(PC_{f, {\cal T}}\). In several attacks~\cite{Johansson_Meier_Muller06,Hell_Johansson06,achterbahn128}, it was assumed that the piling-up lemma~\cite{Matsui93} holds, {\em i.e.}
\[\E(PC_{f, {\cal T}}) = \left[\E(f\oplus g)\right]^{2^s}.\]
But it clearly appears that this result does not apply since the
terms \(f(x_1(t+\tau), \ldots, x_n(t+\tau))\) for the different values
of \(\tau \in {\cal T}\) are not independent. Actually, Naya-Plasencia~\cite{Naya07} and Hell and Johansson \cite{Hell_Johansson07} have independently pointed out that the so-called {\em piling-up approximation}~\cite{Kukorelly} is far from being valid in some cases. 

For instance, the \(11\)-variable Boolean function used in Achterbahn-80 is \(6\)-resilient. An exhaustive search for the initial states of \(\bx_1\) and \(\bx_2\) and a decimation by \(T_7\) enable the attacker to use parity-check relations for \(f^\prime = f + x_1 + x_2 + x_7\), which is \(3\)-resilient. Then, the quadratic approximation 
\[g=x_3x_{10}+x_4x_9 \mbox{ with }\E(f^\prime \oplus g) = 2^{-5}\]
has been considered, corresponding to the set
\[{\cal T}=\{c_1 T_3 T_{10} + c_2 T_4T_9, \; c_1, c_2 \in \{0,1\}\}.\]
It has been deduced that the bias of \(PC_{f^\prime, {\cal T}}\) was \((2^{-5})^4 = 2^{-20}\), leading to an infeasible attack which exceeds the keystream length limitation~\cite{achterbahn128}: the data complexity must be at least \(2^{40}\) and must be multiplied by \(T_7=2^{28}\). But, Naya-Plasencia in~\cite{Naya07} used another approximation, namely
\[g= x_3+x_{10}+ x_4+x_9 \mbox{ with }\E(f^\prime\oplus g) = 2^{-3}.\]
This linear approximation leads to 
\(\E(PC_{f, {\cal T}}) = 2^{-12}\)
for the same set \({\cal T}\), and to a feasible attack with an overall data complexity close to \(2^{52}\) (see~\cite{Naya07} for a precise estimation of the complexity). 

From this concrete example, it clearly appears that estimating the bias of \(PC_{f, {\cal T}}\) may be a difficult problem. This issue has been raised in~\cite{Naya07, Gottfert_Gammel07} which have identified some cases where the piling-up approximation holds. However, since these equality cases are quite rare, a much more extensive study is needed in order to evaluate the resistance of such keystream generators to distinguishing attacks. In this paper, we first emphasize that, even if most attacks based on parity-check relations use an explicit correspondence between the set \({\cal T}\) and an approximation~\(g\) of~\(f\) depending on \(k\)~variables, the bias of \(PC_{f, {\cal T}}\) does not depend on this approximation. Most notably, we show in the next section that the piling-up lemma applied to any approximation~\(g\) compatible with~\({\cal T}\) provides a lower bound on \(\E(PC_{f, {\cal T}})\). Then, Section~\ref{section3} gives two exact expressions for \(\E(PC_{f, {\cal T}})\), one involving the biases of some restrictions of \(f\), and the other one by means of its Walsh coefficients. These expressions lead to an algorithm for computing the bias of a parity-check relation with a much lower complexity than the usual approach, and they also provide some simple formulae for this bias in some particular cases which are commonly used in cryptography, especially when \(f\) is a plateaued function.

\section{A lower bound on the bias of parity-check relations}\label{section2}

However, we can prove that the piling-up approximation
provides a lower bound on the bias of \(PC_{f, {\cal T}}\).

\begin{theorem}\label{th1}
Let \(\bx_1, \ldots, \bx_n\) be \(n\) sequences with least periods \(T_1, \ldots, T_n\), \(f\) a Boolean function of \(n\)~variables and \(\bs = f(\bx_1, \ldots, \bx_n)\). Let
\[{\cal T}=\{\sum_{i=1}^s c_i M_i, \;\; c_i \in \{0,1\}\}\]
where \(M_i = q_i {\rm lcm}(T_{\ell_i+1}, \ldots, T_{\ell_{i+1}})\) with \(q_i > 0\), \(\ell_1=0\) and \(\ell_{s+1}=k\).
Then, for any Boolean function \(g\) of \(k\)~variables of the form
\begin{equation}
\label{eq: g}
g(x_1, \ldots, x_k) = \sum_{i=1}^s g_i(x_{\ell_i+1}, \ldots, x_{\ell_{i+1}})
\end{equation}
where each \(g_i\) is a Boolean function of \((\ell_{i+1}-\ell_i)\)~variables, we have
\[\E(PC_{f, {\cal T}}) \geq \left[\E(f \oplus g)\right]^{2^s}.\]
\end{theorem}

The keypoint in the previous theorem is that \(\E(f \oplus g)\) provides a lower bound on the bias on the parity-check relation for any choice of the approximation~\(g\) of the form~(\ref{eq: g}). The linear approximation of~\(f\) by the sum of the first \(k\)~input variables is usually considered, but any linear approximation involving these variables can be chosen, as stated in the next corollary. 
In the following, 
for any \(\alpha \in {\bf F}_2^n\),
\(\varphi_\alpha\) denotes the linear function of \(n\)~variables: \(x \mapsto
\alpha \cdot x\), where \(x \cdot y\) is the usual scalar product.
\begin{corollary}\label{coro1}
With the notation of Theorem~\ref{th1}, we have
\[\E(PC_{f, {\cal T}}) \geq \max_{\alpha \in V_k} \left[\E(f \oplus \varphi_{\alpha})\right]^{2^{s}}\]
where \(V_k\) is the subspace spanned by the first \(k\) basis vectors.
\end{corollary}
It is worth noticing that this corollary leads to a lower bound on the bias of the parity check relation even if the functions \(f\) and \(x \mapsto x_1 \oplus \ldots \oplus x_k\) are not correlated ({\em i.e.}, if the Walsh coefficient of \(f\) at point \(1_{k}\) vanishes, where the first \(k\)~coordinates of \(1_{k}\) are~\(1\) and the other~\((n-k)\) are zero). This is the first known result in such a situation; the impossibility of deducing any estimation of the bias of the relation in such cases has been stressed in Example~1 in~\cite{Gottfert_Gammel07}.

However, some other approximations~\(g\) with a higher degree may lead to a better bound. But, since any Boolean function is completely determined by its Walsh transform, {\em i.e.} by the biases of all its linear approximations, it appears that \(\E(PC_{f, {\cal T}})\) can be computed from the biases of the linear approximations of~\(f\) only.

\section{Exact formulae for the bias of the parity-check relation}\label{section3}

In some situations, especially when the designer of a generator has to guarantee that the system resists distinguishing attacks, the previous lower bound on the bias of a parity-check relation is not sufficient, and its exact value must be computed. However, since a parity-check relation with \(2^s\)~terms involves \(n 2^s\)~variables where \(n\) is the number of variables of~\(f\), computing its bias requires \(2^{n2^s}\)~evaluations of~\(f\), which is out of reach in many practical situations. For instance, Achterbahn-128 uses a combining function~\(f\) of \(13\)~variables, and the biases of parity-check relations with \(8\)~terms ({\em i.e.} with \(s=3\)) must be estimated; this requires \(2^{104}\)~operations.
Here, we give two exact expressions of the bias of a parity-check relation, which can be computed with much fewer operations, {\em e.g.} with \(2^{43}\)~evaluations of \(f\) in the previous case. The first expression makes use of the biases of the restrictions of \(f\) when its first \(k\)~inputs are fixed; the second one, which is related to a theorem due to Nyberg~\cite{Nyberg01}, is based on the Walsh coefficients of the combining function. A similar technique is also used in another context in~\cite{LV04}.

\subsection{Expression by means of the restrictions of \(f\)}
\begin{definition}
Let \(f\) be a Boolean function of \(n\)~variables and let \(V_k\) and \(V_{n-k}\) be two subspaces such that \(V_k \times V_{n-k} = {\bf F}_2^n\) and \(\dim(V_k)=k\). Then, the restriction of \(f\) to the affine subspace \(a + V_{n-k}\), \(a \in V_k\), denoted by \(f_{a+V_{n-k}}\), is the Boolean function of \((n-k)\) variables defined by
\[f_{a+V_{n-k}}: x \in V_{n-k} \mapsto f(x+ a).\]
\end{definition}

Now, for computing the exact value of \(\E(PC_{f, {\cal T}})\), we
decompose \(PC_{f, {\cal T}}\) according to the values of the first
\(k\)~variables in~\(f\) since the other \((n-k)\)~sequences
\(\bx_i\), \(k+1 \leq i \leq n\), are supposed to be such that \(x_i(t + \tau)\) is
statistically independent from \(x_i(t)\) for any \(\tau \in {\cal T}\).
Amongst the \(k 2^s\) variables \(x_i(t+\tau)\), \(1 \leq i \leq k\)
and \(\tau \in {\cal T}\), we can easily see that each variable is
repeated once. Indeed, for \(j\) such that \(\ell_i < j \leq
\ell_{i+1}\) we have
\(x_j(t + \tau) = x_j(t+\tau^\prime)\)
if and only if \(|\tau - \tau^\prime| = M_i\).

It follows that the values of \(x_j(t+\tau)\), \(1 \leq j \leq k\) and \(\tau \in {\cal T}\) are determined by a \(k 2^{s-1}\)-bit word \(\alpha\). Let us split \(\alpha\) into \(k\) words \((\alpha_1, \ldots, \alpha_k)\) of \(2^{s-1}\)~bits.
We use the correspondence between the values of \(\tau = \sum_{i=1}^s c_i M_i\) in \({\cal T}\) and the integers \(c\), \(0 \leq c \leq 2^s-1\) defined by \(c = \sum_{i=1}^{s} c_i 2^{i-1}\). Then, the value of the \(k\)-bit word \((x_1(t+\tau), \ldots, x_k(t+\tau))\) is equal to
\(\chi(c, \alpha) = (\chi_1(c,\alpha), \ldots, \chi_k(c,\alpha))\) 
where, for any \(j\) such that \(\ell_i < j \leq \ell_{i+1}\), we have 
\[\chi_j(c,\alpha) = \left\{ \begin{array}{ll}
\chi_j(c-2^i,\alpha) & \mbox{ if } c_i \neq 0\\
\alpha_{j, 2^i q+r} & \mbox{ if } c = 2^{i+1}q +r, r< 2^i.
\end{array}\right.\]
Clearly, if \(c_i \neq 0\), we have that \(c\) and \(c^\prime = c-2^i\) correspond to a pair \((\tau, \tau^\prime)\) with \(\tau - \tau^\prime = M_i\). Since \(M_i\) is a period of \(\bx_j\), we deduce that \(\chi_j(c, \alpha) = \chi_j(c^\prime, \alpha)\).

If \(c_i = 0\), the corresponding value of \(x_j(t + \tau)\) is statistically independent from the previous ones and must be defined by a bit of \(\alpha\) which has not been used for smaller values of~\(c\). The number of bits of \(\alpha_j\) which has been used for previous vectors \(\chi_j(c^\prime, \alpha)\) for \(c^\prime < 2^{i+1}q\) is \(2^iq\) since the set \(\{0, \ldots, 2^{i+1}q-1\}\) is composed of \(2^iq\) pairs of the form \((c^\prime, c^\prime+2^i)\) with \(c^\prime_i =0\). Moreover, all \(c^\prime\) in \(\{2^{i+1}q, \ldots, 2^{i+1}q+ r-1\}\) satisfy \(c^\prime_i=0\) because \(r < 2^i\). Therefore, exactly \((2^{i}q+r-1)\) bits of \(\alpha_j\) have been used for \(\chi_{j}(c^\prime, \alpha)\), \(c^\prime < 2^{i+1}q+r\).

\begin{example}
Let us consider a set \({\cal T}\) composed of \(2^3\) elements which involve the periods of \(4\)~sequences:
\[{\cal T}=\big\{c_1 T_1T_2 + c_2 T_3 + c_3 T_4, \; c_1, c_2, c_3 \in \{0.1\}\big\}.\] Then, the \(4\)-bit words \(\chi(c, \alpha)\), \(0 \leq c < 8\), are defined by the \(16\)-bit word \(\alpha\) as follows, where the bold elements correspond to those which have already been used for a smaller value of~\(c\):
\begin{eqnarray*}
\chi(0, \alpha) = (\alpha_{00} \alpha_{10} \alpha_{20} \alpha_{30}) &  & 
\chi(4, \alpha) = (\alpha_{02} \alpha_{12} \alpha_{22} {\bf \alpha_{30}}) \\
\chi(1, \alpha) = ({\bf \alpha_{00}} {\bf \alpha_{10}} \alpha_{21} \alpha_{31}) &  & 
\chi(5, \alpha) = ({\bf \alpha_{02}} {\bf \alpha_{12}} \alpha_{23} {\bf \alpha_{31}}) \\
\chi(2, \alpha) = (\alpha_{01} \alpha_{11} {\bf \alpha_{20}} \alpha_{32}) &  & 
\chi(6, \alpha) = (\alpha_{03} \alpha_{13} {\bf \alpha_{22}} {\bf \alpha_{32}}) \\
\chi(3, \alpha) = ({\bf \alpha_{01}} {\bf \alpha_{11}} {\bf \alpha_{21}} a_{33}) &  & 
\chi(7, \alpha) = ({\bf \alpha_{03}} {\bf \alpha_{13}} {\bf \alpha_{23}} {\bf \alpha_{33}}) 
\end{eqnarray*}
\end{example}

The definition of \(\chi(c, \alpha)\) enables us to express the bias of \(PC_{f, {\cal T}}\) by means of the biases of the restrictions of \(f\) to all cosets of the subspace~\(V_{n-k}\) spanned by the last \((n-k)\)~basis vectors.
\begin{theorem}
Let \(\bx_1, \ldots, \bx_n\) be \(n\) sequences with least periods \(T_1, \ldots, T_n\), \(f\) a Boolean function of \(n\)~variables and \(\bs = f(\bx_1, \ldots, \bx_n)\). Let
\[{\cal T}=\{\sum_{i=1}^s c_i M_i, \;\; c_i \in \{0,1\}\}\]
where \(M_i = q_i {\rm lcm}(T_{\ell_i+1}, \ldots, T_{\ell_{i+1}})\) with \(q_i > 0\), \(\ell_1=0\) and \(\ell_{s+1}=k\).
Assume that \({\cal T}\) does not contain any multiple of \(T_j\), for any \(k < j \leq n\).
Let \(V_{n-k}\) be the subspace spanned by the last \((n-k)\) basis vectors.
Then, we have
\[\E(PC_{f, {\cal T}}) = \frac{1}{2^{k2^{s-1}}}\sum_{\alpha \in {\bf F}_2^{k2^{s-1}}} \prod_{c=0}^{2^s-1} \E(f_{\chi(c,\alpha)+V_{n-k}}).\]
\end{theorem}
\begin{IEEEproof}
\begin{eqnarray*}
\Pr[PC_{f, {\cal T}}(t)=0] & = & \frac{1}{2^{k2^{s-1}}}\sum_{\alpha \in {\bf F}_2^{k2^{s-1}}} \Pr[PC_{f, {\cal T}}(t)=0 | \\
& & (x_1(t+\tau), \ldots, x_k(t+\tau)) = \chi(c, \alpha)].
\end{eqnarray*}
When the values of the first \(k\) input variables in every term of \(PC_{f, {\cal T}}\) are fixed, the piling-up lemma can be applied since the remaining \((n-k)2^s\) variables are statistically independent. The reason is that \(\tau\) is not a multiple of the period~\(T_i\), for any \(k < i \leq n\). Then, we deduce that the term corresponding to~\(\alpha\) in the previous sum equals
\begin{eqnarray*}
&&\frac{1}{2} \left[ 1 + \prod_{\tau \in {\cal T}} \E(f(x(t+\tau), y(t+\tau)) | x(t+\tau) = \chi(c, \alpha)) \right] = \\
&&\frac{1}{2} \left[ 1 + \prod_{c=0}^{2^s-1} \E(f_{\chi(c,\alpha)+V_{n-k}}) \right].
\end{eqnarray*}
We then deduce that
\[\Pr[PC_{f, {\cal T}}(t)=0] = \frac{1}{2} \!\! \left[ 1 \!\!+ \!\!\frac{1}{2^{k2^{s-1}}}\!\!\sum_{\alpha \in {\bf F}_2^{k2^{s-1}}}\prod_{c=0}^{2^s-1} \E(f_{\chi(c,\alpha)+V_{n-k}}) \right].\]
\end{IEEEproof}

This result provides an algorithm for computing the exact value of \(\E(PC_{f, {\cal T}})\).  
The precomputation step consists in computing and storing in a table the \(2^k\)~values of \(\E(f_{a + V_{n-k}}) = \frac{1}{2^k} \sum_{y \in V{n-k}} (-1)^{f(a+y)}\), for all \(a \in V_k\). This step requires \(2^n\) evaluations of~\(f\). Then, computing the bias of the parity-check relation needs to compute, for all \(\alpha \in {\bf F}_2^{k2^{s-1}}\), the product of \(2^s\) precomputed values whose indexes are given by \(\chi(c, \alpha)\), for \(0 \leq c < 2^s\). This requires \(2^{k2^{s-1}} \times 2^s\) operations over integers.
This leads to an overall complexity of \(2^{k2^{s-1}+s} + 2^n\) which is much lower than the complexity of the trivial computation, \(2^{n2^s}\) evaluations of \(f\). For instance, the \(13\)-variable function in Achterbahn-128 is \(8\)-resilient. Estimating the bias of a parity-check relation involving \(10\)~input variables with \(8\)~terms ({\em i.e.} with \(s=3\)) then requires \(2^{43}\)~operations.

\subsection{Expression by means of the Walsh coefficients of \(f\)}
A similar exact expression for the bias of \(\E(PC_{f, {\cal T}})\) can be obtained from the Walsh coefficients of \(f\), {\em i.e.} from all biases 
\(\E(f +\varphi_{a}), a \in V_k\) where \(V_k\) is the subspace spanned by the first \(k\) basis vectors.
\begin{theorem}\label{th3}
Let \(\bx_1, \ldots, \bx_n\) be \(n\) sequences with least periods \(T_1, \ldots, T_n\), \(f\) a Boolean function of \(n\)~variables and \(\bs = f(\bx_1, \ldots, \bx_n)\). Let
\[{\cal T}=\big \{\sum_{i=1}^s c_i M_i, \;\; c_i \in \{0,1\} \big \}\]
where \(M_i = q_i {\rm lcm}(T_{\ell_i+1}, \ldots, T_{\ell_{i+1}})\) with \(q_i > 0\), \(\ell_1=0\) and \(\ell_{s+1}=k\).
Assume that \({\cal T}\) does not contain any multiple of \(T_j\), for any \(k < j \leq n\).
Then, we have
\[\E(PC_{f, {\cal T}}) = \sum_{\alpha \in {\bf F}_2^{k2^{s-1}}} \prod_{c=0}^{2^s-1} \E(f + \varphi_{\chi(c, \alpha)}).\]
\end{theorem}
This expression leads to an algorithm for computing the bias which is very similar to the one based on the biases of the restrictions of~\(f\). But, we need to precompute and to store the Walsh coefficients of~\(f\) corresponding to all elements in~\(V_k\).

\section{Computing the bias in some particular cases}

As a direct corollary of Theorem~\ref{th3}, we obtain the following theorem. It shows that equality holds in Corollary~\ref{coro1} when, amongst all linear functions depending on the \(k\)~variables involved in \({\cal T}\), a single one corresponds to a biased approximation of~\(f\). With this theorem, we recover the value of the bias of a parity-check relation involving the periods of \(k\)~input sequences when the resiliency order of \(f\) is equal to~\((k-1)\). This particular case of our theorem corresponds to the case identified in~\cite{Naya07,Gottfert_Gammel07} where the piling-up approximation holds.

\begin{theorem}\label{th4}
With the notation of Theorem~\ref{th3}, suppose that there exists a single linear function \(\varphi_a\) with \(a \in V_k\) such that \(\E(f + \varphi_a) \neq 0\). Then, we have
\[\E(PC_{f, {\cal T}}) = \left[\E(f + \varphi_{a})\right]^{2^s}.\]
In particular, if \(f\) is \((k-1)\)-resilient, then
\[\E(PC_{f, {\cal T}}) = \left[\E(f + \varphi_{1_k})\right]^{2^s}.\]
where \(1_{k}\) is the \(n\)-bit word whose first \(k\)~coordinates are equal to~\(1\) and the other ones are equal to~\(0\).
\end{theorem}

For a \(t\)-resilient function, the bias of a parity-check relation involving any \((t+1)\)~inputs is given by Theorem~\ref{th4} but, as pointed out in~\cite{Gottfert_Gammel07}, this result does not hold anymore when \({\cal T}\) involves \((t+2)\)~sequences. However, this case can be treated when the function~\(f\) is plateaued~\cite{Zheng_Zhang99}, {\em i.e.} when all values taken by its Walsh transform belong to \(\{0, \pm W\}\) for some \(W\). Note that both combining functions in Achterbahn-80 and in Achterbahn-128 are plateaued. 
 
\begin{theorem}
With the notation and hypotheses of Theorem~\ref{th3}, suppose that \(f\) is \((k-2)\)-resilient and plateaued, {\em i.e.} \(\E(f + \varphi_a) \in \{0, \pm \varepsilon\}\) for all \(a \in {\bf F}_2^n\). Let 
\[{\cal A} = \{ a \in V_k, \E(f + \varphi_a) \neq 0\}.\]
Then, 
\[\E(PC_{f, {\cal T}}) \leq |A|^{2^{s-1}} \varepsilon^{2^s}.\]
Moreover, equality holds if and only if there exists \(i\), \(1 \leq i \leq s\), such that \(M_i\) is a period of all sequences \(\bx_j\) for all \(j\) in \(\cup_{a \in {\cal A}} {\rm supp}(1_k \oplus a)\).
\end{theorem}

\section*{Acknowledgment}
This work was supported in part by the French Agence Nationale de la Recherche under Contract ANR-06-SETI-013-RAPIDE.

\nocite{GGK06,achterbahn2,achterbahn3}
\bibliographystyle{IEEEtran.bst}

\end{document}